\documentclass[aps,twocolumn,showpacs,superscriptaddress, 
amsmath,amssymb]{revtex4} 
 
\usepackage{graphicx} 
 
\begin{document} 
 
\title{Control of bound-pair transport by periodic driving} 
\author{K. Kudo} 
\affiliation{Division of Advanced Sciences, Ochadai Academic Production, 
Ochanomizu University, 2-1-1 Ohtsuka, Bunkyo-ku, Tokyo 112-8610, Japan} 
\author{T. Boness} 
\author{T.S. Monteiro} 
\affiliation{Department of Physics and Astronomy, 
University College London, Gower Street, London WC1E 6BT, United Kingdom} 
 
\date{\today} 
\begin{abstract} 
We investigate  the effect of periodic driving by an external field on systems with attractive pairing interactions. These include
 spin systems (like the ferromagnetic XXZ model) as well as ultracold
 fermionic atoms described by the attractive Hubbard model. We show that
 a well-known phenomenon seen in periodically driven 
systems---the renormalization of the exchange coupling strength---acts  
selectively on bound-pairs of spins/atoms, relative to  magnon/bare atom states. Thus one can control the direction and speed of transport of bound-pair relative to  
magnon/unpaired atom states, and thus coherently achieve spatial separation of these 
components. Applications to recent experiments on transport with fermionic atoms in optical lattices
 which consist of mixtures of bound-pairs and bare atoms are discussed. 
\end{abstract} 
\pacs{75.10.Pq, 67.85.Hj, 05.60.Gg, 03.75.Lm} 
\maketitle 

\section{Introduction} 
 
The dynamics of quantum spin transport and in particular spin correlations are of much current interest 
 in the field of quantum information \cite{Amico}.  
Cold-atoms in optical lattices provide clean realizations of a range of many-body Hamiltonians, including
well-known spin systems. There is also enormous interest in pairing
phenomena, motivated by many ground-breaking experiments with ultracold fermionic atoms in optical
lattices \cite{Fermi}.
 
Neglecting particle interactions, many simple many-body Hamiltonians, such as Heisenberg,
 Hubbard or Bose-Hubbard, 
comprise a hopping term $-JH_{\rm h}$ characterized by an exchange coupling strength $J$. If such a 
system is subjected to an additional 
spatially-linear, but oscillating field, we have a total time-dependent Hamiltonian: 
\begin{equation} 
H(t)= -J H_{\rm h} + B \sin \omega t \sum_{n=1}^{N} n
\sigma_{n}^z / 2.
\label{Ht} 
\end{equation} 
For these, as well as other analogous driven systems, there are
regimes where the exchange strength takes an effective, renormalized, value \cite{DL,CDT,Grifoni,Holt}: 
\begin{equation} 
J_{\rm eff}= J \mathcal{J}_0(\frac{B}{\omega}), 
\label{renorm} 
\end{equation} 
where $\mathcal{J}_0$ denotes an ordinary Bessel function.
The inter-site transport is completely suppressed 
 at values of $\frac{B}{\omega}=2.4$, $5.52$, ...   
corresponding to $\mathcal{J}_0(\frac{B}{\omega})=0$.
The oscillating potential in Eq.(\ref{Ht}) can be implemented 
with ultracold atoms in shaken optical lattices; thus   
Eq.(\ref{renorm}) was recently demonstrated  experimentally \cite{Arimondo,Kierig} by mapping $|J_{\rm eff}|$
as a function of the ratio $\frac{B}{\omega}$.

 In the high frequency regime, where $B, \omega  \gg J$, the suppression is often termed Coherent Destruction of Tunneling (CDT) \cite{CDT,Grifoni}.
 The behavior of the underlying quantum spectrum 
 and the application of CDT to controlling the superfluid-Mott Insulator
 transition in atomic systems were investigated theoretically in 
Ref.~\cite{Holthaus,Creffield}.

 However, this CDT is also closely related to a phenomenon  
 termed Dynamic Localization (DL)---identified even earlier in 
1986 \cite{DL}---which also impedes transport
 in periodically driven systems at $\mathcal{J}_0(\frac{B}{\omega}) \simeq 0$. It too has been
 theoretically investigated in atomic systems  \cite{DL1,Eckardt09}.
 The precise relationship between CDT and DL remains of much interest: a recent theoretical analysis is provided
by  Ref.~\cite{Saito}, but in brief, the CDT mechanism is associated
 with high-frequency driving ($\omega \gg J$) and complete suppression
 of hopping. For CDT, particles remain completely frozen at their
 original sites even for a 2-site system: CDT was initially identified
 in driven double-well systems \cite{CDT,Grifoni}. CDT can persist even
 in the presence of some inter-particle interactions
 \cite{Holthaus,Creffield}. DL, on the other hand, entails a less
 complete suppression of transport: for 
$\mathcal{J}_0(\frac{B}{\omega}) \simeq 0$, 
the single-particle wavepacket position may oscillate, but
 the particle returns periodically to its original position. 
  DL is associated with lower frequency driving, negligible particle
 interactions and the large $N$ limit \cite{Saito}.

Here we investigate for the first time periodic-driving for systems with
a bound-pair component. In the absence of driving, transport for
fermionic systems with attractive pairing  
interactions has been previously investigated in, e.g., ferromagnetic
XXZ spin models \cite{Santos03} and even experimentally with ultracold
atoms which are a mixture of 
bound-pairs and bare atoms, in a regime of the attractive Hubbard model
\cite{BP}. 
 Our main finding is that the DL regime (but not CDT) can  provide a mechanism 
 to {\em spatially separate} the paired and unpaired fraction in such spin or cold atoms systems. This is because the driving can
 generate a hitherto unnoticed mechanism of directed motion; 
in contrast, previous studies to date, such as Refs.~\cite{Arimondo,Creffield,Holthaus}, probed only
$|J_{\rm eff}|$, i.e., the presence or absence 
of diffusive expansion of an atomic cloud, not global transport.

We note that both the DL and CDT transport-suppression mechanisms are usually analyzed in terms of the stationary states (Floquet states)
of the driven  Hamiltonian: typically, the Floquet eigenstates (i.e.,
their quasienergies) become degenerate or approximately degenerate if
$\mathcal{J}_0(\frac{B}{\omega}) \simeq 0$. 
However, here, in the $N \to \infty$ limit, we analyze the quantum
transport  without any reference to the detailed quantum structure, or
quasienergy near-degeneracies: we obtain the renormalization
Eq.(\ref{renorm}) (as well as the ratchet mechanism) from a
straightforward integration of the classical equations of motion of the
driven Hamiltonian. This entails a somewhat modified perspective, since
now the key division is not between high-frequency CDT
freezing-of-motion and low-frequency DL wavepacket revival. It is rather
between the $N \to \infty$ limit, where both high and 
low frequency suppression of transport arises as a classical phenomenon
(i.e., it is purely an effect in the group velocity), and the small $N$
limit, where a quantal analysis of a few Floquet state degeneracies
remains essential---and where only high frequency driving suppresses
transport. The effect of inter-particle interactions seems difficult to
treat classically. But, below we identify an example where we can
include the effect of certain important interactions  by considering two
limiting effective Hamiltonians. 

The paper is organized as follows:
In Sec. \ref{sec:model}, we introduce our models.
One is a Heisenberg XXZ spin chain; its eigenstates include both magnon-like
 spin-waves (analogous to Bloch waves) and bound-pair states. 
Explicit expressions for both are obtainable via the Bethe ansatz.
The close correspondence with the two-particle attractive Hubbard model,
which also includes both bare atom and bound-pair states, is explained.
The Hubbard model is of particular interest as it is already realized in
current cold atoms experiments. 
In Sec. \ref{sec:CDT}, the quantum CDT mechanism, dominant for low 
$N$ and $\omega \gg J$ is introduced. Its effects are demonstrated for
both unpaired and bound-pair states by exact numerical solution of the
quantum Hamiltonian. In Sec. \ref{sec:DL}, the low-frequency, high-$N$
regime is investigated; control of wavepacket dynamics is shown for the
one and two-particle regimes. The underlying mechanism of directed
motion are obtained from the classical equations of motion.  The
behavior of both bound and unpaired cases are analyzed with two limiting
classical Hamiltonians and are shown to agree with the full quantum
solutions. Conclusions and outlook are given in Sec. \ref{sec:conc}. 

\section{\label{sec:model} Static Models}

In general, inter-particle interactions are important  so the full many-body Hamiltonian takes the form
$J H= J H_{\rm h} + (J\Delta) H_{\rm int}$, where $H_{\rm int}$ is an interaction term characterized
by an interaction strength $U= J\Delta$. For instance,
the well-known spin-$1/2$ 
 Heisenberg XXZ ferromagnetic chain of length $N$, is governed by the Hamiltonian: 
\begin{equation} 
 -J H=-\sum_{n=1}^{N}[\frac{J}{2}(\sigma_{n}^{+}\sigma_{n+1}^{-}+\sigma_{n}^{-}\sigma_{n+1}^{+})+ 
\frac{J\Delta}{4}\sigma_{n}^{z}\sigma_{n+1}^{z}] 
\label{Hhc} 
\end{equation} 
where $n \in [1:N]$ indicates the $n$-th site, and $\Delta$ denotes the anisotropy.
For large $\Delta$ (and low density of excitations), there are two dominant classes of quasi-particle states:
magnon-like and bound-pair states.

The Hamiltonian Eq.(\ref{Hhc}) conserves the number of spin-flips; a  
single excitation represents a 
spin-wave, or magnon, which distributes a single spin-flip throughout the chain. 
The spin-wave eigenstates of Eq.(\ref{Hhc}) are  
$|\kappa \rangle=\frac{1}{\sqrt{N}} \sum_{n=1}^{N} e^{i n\kappa} \,|n\rangle $
(for periodic boundary conditions) where $|n\rangle$ denotes a state with the spin-flip at sites $n$; they obey the  
dispersion relation $E_{\kappa}-E_0=-J\cos \kappa $, where $E_0$ is the ground state.
Single magnon transport is thus analogous to free atoms (Bloch waves) 
in the lowest band of a lattice.
Higher excited states correspond to multiple spin-waves which interact when they 
coincide through (i) an exclusion process (no two spin flips can simultaneously 
occupy the same site) and (ii) an effective attractive interaction induced by the 
$(J\Delta) H_{\rm int}= \sum_n \frac{J\Delta}{4} \sigma^z_{n}\sigma^z_{n+1}$  interaction term. For the long wavelength
processes considered in our key results Figs.\ref{Fig2} and \ref{Fig3}, only (ii) is of
significance.

Now we consider the two-particle case. 
For the two-excitation case, the eigenstates of Eq.(\ref{Hhc}) may be 
expressed, via the Bethe ansatz,   
as spin waves \citep{KarbMull98, Santos03}: 
\begin{equation} 
|\kappa_{1},\kappa_{2}\rangle=A_{\kappa_{1},\kappa_{2}} 
\sum_{ n_{1}<n_{2}\leq N} 
a^{\kappa_{1},\kappa_{2}}_{n_{1},n_{2}}\,|n_{1},n_{2}\rangle,\label{eq:ges} 
\end{equation} 
where $A_{\kappa_{1},\kappa_{2}}$ is a normalization constant and 
$|n_{1},n_{2}\rangle$ indicates a spin-flip at sites $n_1$ and $n_2$. 
 Further details are in Refs.
\cite{Santos03,Boness09} but the eigenstates divide into two distinct classes: 
 (1) Magnon-like {\em scattering states}, 
where the spins move separately, with dispersion relation  
$E_{\kappa_{1},\kappa_{2}}-E_0=J\left(2\Delta-\cos \kappa_{1}-\cos \kappa_{2}\right)$ 
which is a straightforward extension of the singly-excited case and 
(2) {\em Bound-pair states},   for which the 
 probability amplitudes decay exponentially with the separation of the flips. For $\Delta > 0$, and as $N \to \infty$ 
they obey the dispersion relation, 
$E_{\kappa_{1},\kappa_{2}}-E_0=J\Delta-\frac{J}{2\Delta}[1+\cos(\kappa_{1}+\kappa_{2})]$. 
Note that the sum $(\kappa_{1}+\kappa_{2})$ is always real. 
In Ref. \cite{Santos03}  a 
perturbation expansion in spin coupling strength $J$ is employed to produce an effective 
Hamiltonian when $\Delta \gg 1$.  In this approximation the bound state 
amplitudes are
\begin{equation}  
  a^{\kappa_{1},\kappa_{2}}_{n_{1},n_{2}}= \delta_{n_1,n_2-1} e^{i (\kappa_1+\kappa_2) n_1} 
\label{BS}
\end{equation}
and thus the bound states become confined to the nearest 
neighbor (NN) subspace, $\{|n,n+1\rangle\}$; 
their dispersion relation remains unchanged from the above. 
 In ``center of mass'' coordinates $2\kappa= \kappa_1+\kappa_2$,  the
 bound states are  
given by $|\kappa \rangle=A_{\kappa} \sum_{n=1}^{N}e^{2in \kappa } \,|n,n+1\rangle$  
and are of analogous form to the 
single magnon solution. Namely, two initially neighboring 
spin-flips thus hop {\em together} but their propagation speed is slower
relative to a single spin-flip: transport occurs
by a second order effective Hamiltonian and an effective coupling
 $J_{\rm eff}=\frac{J}{2\Delta} < J$.

 The transport is rather analogous to
  that seen in the attractive Hubbard model
$-JH= J \sum_{k,\sigma} (c^{\dagger}_{k\sigma}c_{(k+1)\sigma} +\mbox{\rm H.c.}) + J\Delta \sum_k n_{k\uparrow}n_{k\downarrow}$
 characterized by tunneling amplitude $J$
and interaction $U=J\Delta$; for $U < 0$, local bound-pairs form which tunnel by a second-order process,
with amplitude of order $\sim J^2/U \sim J/\Delta$
 akin to the spin system. This situation was investigated experimentally using fermionic atoms for $U<0$ in Ref.~\cite{BP};
 note of course that the atomic local bound-pairs (BPs) correspond to a pair of atoms
of opposite spin occupying the same lattice site, while the spin BPs occupy adjacent sites.

\section{\label{sec:CDT} Quantum dynamics: High frequency driving and coherent destruction of tunneling}

\begin{figure}[tb] 
\includegraphics[width=3.5in]{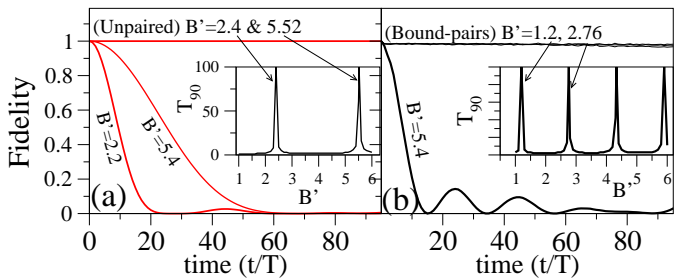} 
\caption{(color online) CDT demonstrated in a XXZ chain with $N=20$ and large 
 $\Delta=8$. For $\Delta \gg 1$,  initial spin-states of
 adjacent spin flips $|\psi(t=0)\rangle = |n, n+1\rangle$ overlap only with bound-pair states;
initial states with well-separated spin-flips 
$|\psi(t=0)\rangle = |n_1,n_2\rangle =|5,15\rangle$
 overlap only with magnon-like scattering states.
The fidelity $F(t)=|\langle\psi(t=0) |\psi(t)\rangle |^2$ is shown as a function of time.
Insets plot $T_{90}$, the time taken for the fidelity to fall to $0.9$, so show the
CDT resonances. {\bf (a)} $J'=1/8$. CDT for magnon states. Spins freeze
 at zeros of $\mathcal{J}_0(B')$ ($B' \simeq 2.4, 5.5 ...$)  but delocalize elsewhere.
{\bf (b)} $\frac{J'}{2\Delta}=1/8$. CDT for bound-pairs. Adjacent spins freeze at
 fields for which $\mathcal{J}_0(2B')=0$
($B'\simeq 1.2,2.7 ...$)} 
\label{Fig1} 
\end{figure} 

  We now consider the regime with $B, \omega  \gg J$ which can lead to
coherent destruction of tunneling. Using  units 
 of scaled time $t'= \omega t $,  the quantum dynamics generated by Eq.(\ref{Ht}) 
depends only on two global parameters $J'=J/\omega$ and  
$B'=B/\omega$ (for a given $\Delta$). The high-frequency 
 driving condition for CDT thus becomes $J'\ll 1$.  
We follow the usual procedure for justifying CDT \cite{Holthaus,Creffield}: for 
 $B' \gg J'$, we initially neglect $H_{\rm h}$ and set
 $H(t)= B'\sin  t' \sum_{n=1}^{N} n\sigma_{n}^{z}/2$. 
 
 Then, the Schr{\"o}dinger equation has the following time-dependent solutions: 
\begin{equation} 
|n_1,n_2,t \rangle= e^{[iB' (n_1+n_2) \cos t]} |n_1,n_2\rangle, 
\label{CDT} 
\end{equation} 
to within a constant phase term $e^{[iB'\varphi \cos t]}$ where
$\varphi=\sum_{n=1}^N n =N(N+1)/4$. 
The stationary states of periodically driven systems (Floquet states) have a Brillouin zone type structure 
\cite{Holthaus,Creffield} and the above correspond to a family of  solutions 
$|n_1,n_2,m,t \rangle =  |n_1,n_2,t\rangle e^{imt}$ which are periodic in time.  
 The multi-photon band index $m$ is used to label states which differ only by  
 the absorption or emission of $m$ quanta of energy $E_m=m\hbar \omega$. 
The next step is to 
treat the spin-exchange as a perturbation, in the extended Hilbert space where time appears as 
an extra coordinate. The  matrix
elements of $H_{\rm h}$ are simply 
\begin{eqnarray} 
\langle \langle n_1',n_2',m',t | J H_{\rm h} |n_1,n_2,m,t \rangle\rangle \simeq \delta_{mm'} \nonumber \\ 
        J \mathcal{J}_0\left[B'(n_1'-n_1)+ B'(n_2'-n_2)\right] \langle n_1',n_2' | 
	 H_{\rm h} |n_1,n_2\rangle. 
\label{mat} 
\end{eqnarray} 
 
Here $\langle \langle . | . \rangle\rangle= \frac{1}{T} \int_0^T \langle .|. \rangle$ 
denotes a scalar product in the extended space \cite{Holthaus} and $\langle .|. \rangle$ denotes 
the scalar product in position coordinates. It is easily seen that in general, the matrix elements 
$\langle n_1',n_2' | H_{\rm h} |n_1,n_2\rangle= 
\left[\delta_{n_1',n_1\pm1}\delta_{n_2',n_2}+\delta_{n_2',n_2\pm1}\delta_{n_1',n_1}\right] 
\langle n_1',n_2' | H_{\rm h} |n_1,n_2\rangle $.   
 In other words, they involve hopping of single spins only. 
 Thus, we have
\begin{equation}
 J \mathcal{J}_0\left[B'(n_1'-n_1)+ B'(n_2'-n_2)\right]=J \mathcal{J}_0(B')
\end{equation}
as in Eq.(\ref{renorm}). This also corresponds to the situation explored in the cold atoms
experiments \cite{Arimondo,Kierig}. A more careful analysis \cite{Holthaus} reveals terms
coupling off-diagonal in the photon band index $m$. If these off-diagonal terms are significant,
the above analysis is no longer valid; however, it has been shown that provided the energy scale 
for the band separation far exceeds the intra-band energy scale (i.e.,
$\omega \gg J$), these 
couplings are negligible. Thus within the above framework, the Eq.(\ref{renorm}) renormalization
holds for high-frequency driving.

However, the {\em bound pairs}, for $\Delta \gg 1$,
evolve under an effective second-order Hamiltonian $H_{\rm eff}$ \cite{Santos03} 
 which results in 
the two spins hopping together. When $n_2 = n_1 + 1$, for instance,
the argument of the Bessel function becomes 
\begin{equation}
 B'(n_1'-n_1)+ B'[n_1'+1 - (n_1+1)] = 2B'(n_1'-n_1).
\end{equation}
Thus, the transport 
is then determined by matrix elements 
 $\frac{J}{2\Delta}\mathcal{J}_0(2 B')\langle n,n+1|H_{\rm eff}|n\pm 1,n+1\pm 1\rangle$:
 note the doubling of the argument of the Bessel  
function. Since the Bessel is an oscillating function one can then, for example, choose a value 
of  $B'$ for which $J_{\rm eff}\simeq J \mathcal{J}_0(B')$ for the magnon-like
states is positive, while  $J_{\rm eff}\simeq\frac{J}{2\Delta} \mathcal{J}_0(2B')$ for the bound-pairs
 is negative (or zero). 

An arbitrary spin state, in general, has a projection 
on both the magnon-like and the bound-pair eigenstates. For $\Delta \gg 1$, however, one may 
 prepare a good approximation to a pure bound-pair state by simply flipping two adjacent spins of a ferromagnetic 
chain in its ground state (all spins aligned). Conversely, two well-separated spins 
will approximate pure magnon-like states. In a corresponding atomic experiment with Fermionic 
atoms one may either consider unpaired single atoms or a pure bound-pair, as well as a superposition
of these two extremes.

 In Fig.\ref{Fig1} we demonstrate CDT for the two extremes (an initial state 
which is either a pure magnon state or a pure bound-pair for large
$\Delta$). Driving with a field $B'=5.52$  ``freezes'' two initially well-separated spin-flips 
 at their original sites, since $\mathcal{J}_0(B')=0$. 
In the absence of driving, or at values of $B'$ with $\mathcal{J}_0(B') \neq 0$, 
both spins rapidly diffuse along the chain. On the other hand, if the two-flips are initially 
{\em adjacent}, they remain frozen at their positions if 
$\mathcal{J}_0(2B')=0$ and delocalize otherwise. 

Other than for these extremes, CDT (e.g., for initial states which are superpositions of magnon/bound-pairs)
is less effective; while both the magnon-like scattering states 
and bound-pair states  are eigenstates of the Hamiltonian Eq.~(\ref{Hhc}), the respective subspaces
are coupled by the driving: in general, for CDT, the driving strength $B \gg J\Delta$, where
$J(\Delta-2)$ is the energetic separation between the two subspaces.

\section{\label{sec:DL} The large-$N$ regime and Dynamic Localization}

\begin{figure}[tb] 
\includegraphics[width=3.in]{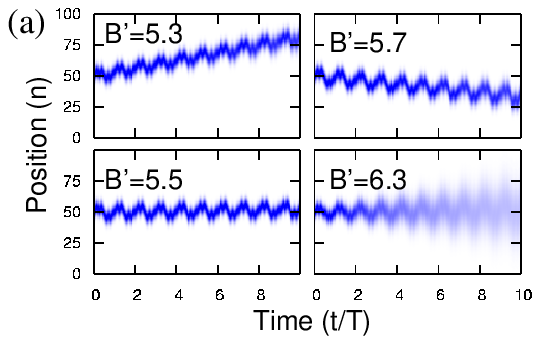} 
\vspace*{10 mm} 
\includegraphics[width=3.in]{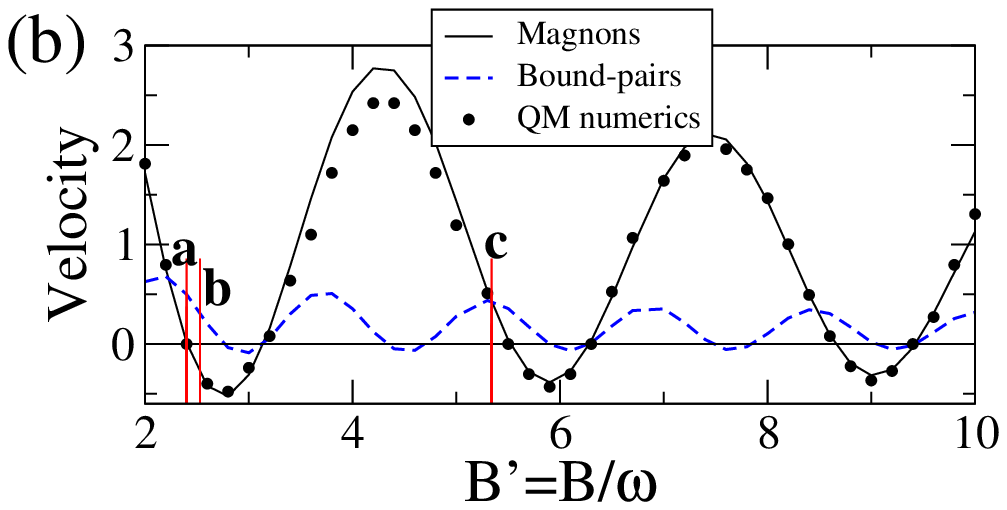} 
\vspace*{-10 mm} 
\caption{(color online) {\bf(a)} Probability distribution $|\psi(n,t)|^2 $ for an initially Gaussian wavepacket
containing a single excitation, near the dynamic localization (DL) regime, for $J'=8$. 
 Two zero-velocity cases are shown: $B'=5.5$ corresponds 
to actual DL, with $\mathcal{J}_0(B')=0$ and suppression of  hopping; while $B'=6.3$ corresponds 
 to $\sin B'=0$, so suppression of directed motion, but not of hopping: the excitation diffuses along the chain. 
{\bf (b)} Average velocity of the wavepackets as a function of $B'$. Numerics were obtained 
 from plots like those in (a). They are well-fitted by the magnon velocity  
Eq.(\ref{magnon}), while the predicted bound-pair behavior corresponds instead to Eq.(\ref{bsv}) 
 shown here for $\Delta=2$. The vertical lines {\bf a,b,c} correspond to the multiply excited packets
shown in Fig. \ref{Fig3}.} 
\label{Fig2} 
\end{figure} 
 
We now investigate the regime $N \to \infty$, usually associated with
Dynamic Localization. 
We can work in the more favorable  regime  $J\Delta \gtrsim B $ where
coupling between bound-pairs and unpaired states is suppressed. 
 We evolve Eq.(\ref{Ht}) for an open chain with $N=100$ sites (for
computational reasons, much larger $N$ is difficult). 
Since our analysis below does not
introduce any high-frequency condition, 
we can also consider low-frequency driving, for which $B',J' \gg 1$. 

We prepare a one-spin flip Gaussian wavepacket, at the center of the
chain. Figure~\ref{Fig2}(a) shows that the spin-packet unexpectedly
moves as a whole, and with little spreading, along the chain, but its
direction of travel {\em reverses} at $B' \simeq 5.5$: for $B'=5.3$ it
moves upwards, for $B'=5.7$ it moves downwards; for $B'=5.5$ it exhibits
behavior characteristic of Dynamic Localization (static, but with
oscillations).   The results for $B'=6.3$ may seem at first sight even
more puzzling: the wavepacket's center of  mass is static, but the
excitation diffuses along the chain. In order to understand  
these results, one notes that Fig.\ref{Fig2}(a) demonstrates a
combination of DL type  renormalization, as well as a (hitherto
unnoticed) type of ratchet (meaning directed motion without bias). 

We note that $N$ plays the role of an effective $\hbar$. 
Thus, in the presence of 
driving and in the $N \to \infty$ limit, one may map the system 
 to an ``image'' classical Hamiltonian:
\begin{equation}
  H(x,p) = -J' \cos p- B'x \sin t'
 \end{equation} 
by mapping site to continuous position $ n \to x$ and $\kappa \to p$. 
Integrating Hamilton's classical equations of motion over one period,
it is easy to see that the distance traveled 
per period $v=\langle x(t+T)-x(t) \rangle /(2\pi)$, where $T=2\pi /\omega$,
for a particle with position $x(t=0)=x_0$ and momentum $p(t=0)=p_0$ is simply: 
\begin{equation} 
 v = J' \mathcal{J}_0(B') \sin (p_0+B').  
\label{magnon} 
\end{equation} 
In effect, this is the center of mass velocity of the wavepacket. It is independent
of initial position $x_0$. The renormalization of the velocity 
already appears, but it is no longer a question of simply renormalizing $J'$ by $\mathcal{J}_0(B')$ because 
there is also a $\sin (p_0+B')$ factor which can also change sign or suppress transport. 
We find, however, that it does provide great advantages in reducing dispersive spreading.

 An initial Gaussian wavepacket, initially 
localized over a finite number of sites $1/\delta_n$, corresponds to localization in momentum 
of $\sim \delta_n$, in the image phasespace. Thus the corresponding momentum distribution is 
$N(p) \sim \exp [-(p_0-\langle p \rangle )^2/\delta_n^2]$. Low-energy spin wavepackets (and cold atom
clouds) correspond to distributions well-localized about zero momentum, i.e.,
$\kappa \sim 0$ thus $\langle p_0 \rangle \simeq 0$. 
The distribution thus samples velocities $ \sim J' \mathcal{J}_0(B') \sin (B' \pm \delta_n)$. Hence provided 
$m\pi \lesssim |B' \pm \delta_n| \lesssim (m+1)\pi $, all momenta $-\delta_n < p_0 < \delta_n$
 correspond to the same direction 
of motion and there is little dispersion. This results in {\em directed motion}. In contrast, 
when $B' \simeq m\pi$ with $m=1,2,\ldots$, trajectories with $p_0 <0$ move in opposite  
direction to those with $p_0> 0$ and the wavepacket, though static, spreads diffusively along the 
chain. In the presence of DL ($B'=5.5$), $\mathcal{J}_0(B')=0$, so there is, of course no transport.
 
In Fig.\ref{Fig2}(b) we see that a calculation of the velocity from the
numerics [displacement 
of the quantum wavepacket per period obtained from a quantum solution of 
Eq.(\ref{Ht})] is in excellent agreement with  Eq.(\ref{magnon}). 
 
\begin{figure}[tb] 
\includegraphics[width=3.1in]{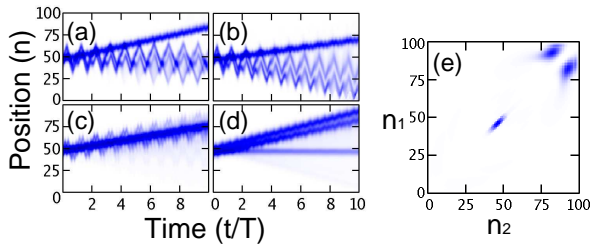} 
\caption{(color online) Dynamics of spin states which are superpositions of 
bound-pair and magnon states, for parameters shown in
Fig.\ref{Fig2}(b).
{\bf(a)}: $J'=10, \Delta=2, B'=2.4$. The magnon states are stopped, 
while  bound-pairs
travel to the end of the spin-chain. {\bf (b)} $J'=10, \Delta=2, B'=2.53$. 
 Bound-pair and magnon states move in opposite direction (bound-pairs
 move upwards).
{\bf(c) } $J'=10, \Delta=2, B'=5.34$. The faster magnon packets are slowed by the
field so magnon and bound-pair speeds become equalized and they now travel together.
 {\bf (d)} $ J'=2, \Delta=5, B'=4.33$. Bound-pairs remain static,
 magnons travel. 
{\bf (e)} Spin-spin correlation,$|\langle n_1,n_2|\psi(t)\rangle|^2$, at
 $t/T=10$ for the spin dynamics of (d).  
The bound-pair component remains stationary, in its initial position at the center, while
the magnon-like components travel to the end of the chain. }
\label{Fig3} 
\end{figure}
  
 Conversely, a single bound-pair state moves under a different 
image Hamiltonian $H= \frac{-J'}{2\Delta}\cos 2P -2B' X \sin t'  $, where  
$2P=p_1+p_2$ and $X=(x_1+x_2)/2$, indicate center of mass coordinates. Thus 
the bound-pair velocity: 
\begin{equation} 
 v_{\rm bs} = \frac{J'}{2\Delta} \mathcal{J}_0(2B') \sin (p_0+2B').  
\label{bsv} 
\end{equation}
For  multiple excitations, we analyze the numerics by assuming that the magnon-like projection 
of the wavepackets obey Eq.(\ref{magnon}), 
while the Hilbert space fraction of  bound-pair states moves under Eq.(\ref{bsv}). We thus assume that 
coupling between the two subspaces is negligible, a reasonable
assumption for $B \ll J\Delta$.

 Figure~\ref{Fig2}(b) (broken line) shows the corresponding velocity. One sees that 
the bound-pair and magnon velocities may even have opposite signs for the same  
driving field. This enables us to steer the two components in opposite directions 
or to stop one and transport the other. These different
situations are demonstrated in Fig.\ref{Fig3}, where the initial state
is the product of two Gaussian wavepackets peaked on $\kappa \simeq 0$ and at
positions $n_1=45$ and $n_2 =50$. The $\kappa \simeq 0$ condition
corresponds to the low-energy distribution---also typical of an
ultracold gas (centered on $p_0 \simeq 0$). 
 There is no special significance in the initial chosen positions
[neither Eq.(\ref{magnon}) nor (\ref{bsv}) depends on initial $x$]
other than choosing an initial state with $n_1$ and $n_2$ close but not adjacent yields an appreciable projection in 
both bound-pair and magnon subspaces. Note also that the extent of the initial distribution should not exceed
that of the chain/lattice, so that center of mass displacement is observable. Equivalently, in a 
Hubbard Hamiltonian, two delocalized atoms with distribution peaked
 one or two sites apart would yield an appreciable probability
 to both form a pair/remain unpaired.

 The directed wavepacket motion is in sharp contrast to the usual motion
seen in studies of CDT/DL transport; also to 
the undriven case: while magnon states diffuse
faster than bound-pair components (by a factor of $2\Delta$) both ultimately simply
delocalize along the length of the chain. Figure~\ref{Fig3}(e)
illustrates the spin correlations for Fig.\ref{Fig3}(d) (for large $\Delta=5$) where the
bound-pair component is immobilized, while the component with separated spins is transmitted.

\section{\label{sec:conc} Conclusions and outlook}
 
In conclusion, we have investigated the transport in a system with
attractive pairing interactions periodically driven by an external
field. The effects of periodic driving act selectively on unpaired and
bound-pair states, which enable us to control separately the relative direction and
speed of each wavepacket of the two states.

One should also consider higher numbers of particles. For low-numbers of excitations (modest
filling factors in the atomic case) the dominant processes are the above. Nevertheless,
the effect of including higher numbers of particles remains an important question.
In the case of high-frequency driving and CDT, this is less of a
concern: e.g., in Ref.~\cite{Creffield}
it was shown that a 2-particle model adequately described the CDT behavior of a full many-body
Hamiltonian. However, for low frequencies, this remains an open question. One of our key findings is
the onset of center of mass motion in the DL regime, which would open new experimental
possibilities, and is well within reach of current techniques. The
experiments of Ref.
\cite{Arimondo} observed the effects of Eq.(\ref{renorm}) in the range $J'\simeq 3 \to 1/30$; for lower
frequencies, the effect was lost due to inter-particle interactions. 
 This range of $J'$ overlaps with our DL wave-packet splitting regime
 [e.g., Fig.\ref{Fig2}(d) 
corresponds to $J'=2$ and $U/J =5$]. In addition, a 
system corresponding to the attractive Hubbard model is somewhat more favorable than the bosonic
atoms used in CDT experiments since there are {\em two} good DL limits 
provided by the fully paired and unpaired extremes, while for the bosonic systems,
low-frequency DL occurs only in the limit of negligible interactions.
A full numerical study of the DL regime (high $N$ and many particles) is numerically
challenging, but future calculations will need to address this question more fully.

 Within the regime of validity of the present work, even for not too large $J\Delta/B$ we see that
the wavepacket  splits cleanly into two components, whose relative speed and direction
may be controlled as in Eqs.(\ref{magnon}) and (\ref{bsv}). This shows that 
 coupling between the magnon and bound-pair subspaces by the external driving is limited.
In fact it is easy to show that, once the wavepackets separate, further interaction is negligible.
If the separation is slow, collisions between the paired and unpaired 
particles may be important; but this itself exemplifies a topic of much current
interest; for example recent 
experiments probed transport of impurity wavepackets accelerating through 
a static cold cloud
by means of a linear potential $V=gx$ due to gravity \cite{Kohl}. The study of
dynamics of bound-pairs moving through a cloud due to the linear oscillating 
lattice potential $V= F x \cos \omega t$
is also within current experimental capabilities and offers new possibilities
for transport in the presence of interactions.

\begin{acknowledgments}
We acknowledge helpful discussions with S. Bose, C. Creffield and 
M. Oberthaler.
This work is partly supported by KAKENHI(21740289).
\end{acknowledgments}

\end{document}